# A stochastic cascade model for Auger-electron emitting radionuclides

Boon Q. Lee[1†], Hooshang Nikjoo[2], Jörgen Ekman[3], Per Jönsson[3], Andrew E. Stuchbery[1] and Tibor Kibédi[1]

[1]Department of Nuclear Physics, Research School of Physics and Engineering, The Australian National University, Canberra, Australia

[2]Department of Oncology-Pathology, Karolinska Institutet, Stockholm, Sweden

[3]Materials Science and Applied Mathematics, Malmö University, Malmö, Sweden

Running Title: MC emission model for Auger emitters.

Keywords: Auger emitters; Auger cascade; $^{123}$I; $^{124}$I; $^{125}$I.

[†]Corresponding author:

Mr. Boon Quan Lee,

Department of Nuclear Physics,

Research School of Physics and Engineering,

The Australian National University,

Canberra, Australia

Tel: (61) 2 6125 0609

Fax: (61) 2 6125 0748

Email: boon.lee@anu.edu.au

**Abstract**

*Purpose:* To benchmark a Monte Carlo model of the Auger cascade that has been developed at the Australian National University (ANU) against the literature data. The model is applicable to any Auger-electron emitting radionuclide with nuclear structure data in the format of the Evaluated Nuclear Structure Data File (ENSDF).

*Materials and methods:* Schönfeld's algorithms and the BrIcc code were incorporated to obtain initial vacancy distributions due to electron capture (EC) and internal conversion (IC), respectively. Atomic transition probabilities were adopted from the Evaluated Atomic Data Library (EADL) for elements with atomic number, $Z$=1-100. Atomic transition energies were evaluated using a relativistic Dirac-Fock method. An energy-restriction protocol was implemented to eliminate energetically forbidden transitions from the simulations.

*Results:* Calculated initial vacancy distributions and average energy spectra of $^{123}I$, $^{124}I$, and $^{125}I$ were compared with the literature data. In addition, simulated kinetic energy spectra and frequency distributions of the number of emitted electrons and photons of the three iodine radionuclides are presented. Some examples of radiation spectra of individual decays are also given.

*Conclusions:* Good agreement with the published data was achieved except for the outer-shell Auger and Coster-Kronig transitions. Nevertheless, the model needs to be compared with experimental data in a future study.

[202 words; target 200]

## Introduction

Auger-electron emitting radionuclides are of great interest for internal radiotherapy due to the very short range of Auger electrons. The subcellular range of Auger electrons is an attractive property for minimizing collateral damage to normal tissues adjacent to the targeted tumour. This unique feature offers some distinct advantages compared to the more commonly used long-range $\beta$-electrons, such as a reduced cross-fire irradiation of non-target healthy cells and a higher ionization density within the immediate vicinity of the decay site, which is generally associated with high biological effectiveness (Behr et al. 2000, Kassis 2004, Buchegger et al. 2006, Nikjoo et al. 2008, Rebischung et al. 2008, Li et al. 2010, Vallis et al. 2014). Auger-emitting radionuclides have shown very promising effects in vitro and in vivo in animal studies over the last decade (Fischer et al. 2008, Chan et al. 2010, Costantini et al. 2010, Koumarianou et al. 2014, Kiess et al. 2015).

Emission spectra of Auger-electron emitting radionuclides are essential for dosimetric calculations to quantify the biological damage delivered to the target (Nikjoo et al. 2008, Bousis et al. 2010, Falzone et al. 2015). In the past three decades several authors published calculated emission spectra of selected radionuclides using either deterministic or Monte Carlo computational methods (Howell 1992, Pomplun 2000, Stepanek 2000, Eckerman and Endo 2007, Nikjoo et al. 2008). There is a large scatter in these published spectra, particularly in the yields and energies of outer-shell transitions, as they were calculated based on different approaches toward evaluation of the Auger cascade. For other less common, but emerging radionuclides such as $^{140}$Nd ($T_{1/2}$=3.4 d) and $^{161}$Tb ($T_{1/2}$=6.9 d), no emission data has yet been made available. $^{140}$Nd is a pure Auger emitter and its short-lived daughter $^{140}$Pr is a positron emitter, which can be used for PET (positron emission tomography) while $^{161}$Tb is a $\beta$/Auger emitter that has been shown to provide an enhanced antitumour effect compared to $^{177}$Lu in an in vivo study (Müller et al. 2014). These radionuclides have more suitable half-lives and better electron-to-photon dose ratios than most of the commonly used Auger emitters. They can now be produced for medical research at Institute of Laue-Langevin (ILL) and CERN-Medical Isotopes Collected from ISOLDE (MEDICIS) (Lehenberger et al. 2011, dos Santos Augusto et al. 2014). Thus, a computational model that can generate the emission spectrum of any Auger emitter in a consistent physics framework is essential for research into Auger-electron targeted radiotherapy.

With that goal in mind, a stochastic Auger-cascade model, which requires only the nuclear structure data of a radionuclide in the format of the Evaluated Nuclear Structure Data File (ENSDF), was recently created to simulate the Auger cascade of elements up to Fermium, $Z$=100 (Lee et al. 2012). The model adopts a strict energy bookkeeping, which is similar to Pomplun's model (2000, 2012), to eliminate any energetically forbidden transitions due to changes in atomic structure during the Auger cascade. The nuclear structure data provided is analyzed in situ to produce the initial vacancy distribution which determines the initial atomic state before the Auger-cascade simulation is begun. The model can produce the full Auger energy spectrum following a nuclear decay of any radioisotope ($Z \leq 100$) or a nuclear reaction, such as $^{157}$Gd(n,$\gamma$)$^{158}$Gd, that can create atomic vacancies. Once validated, the model will open up the opportunities for future microdosimetry studies of uncommon Auger-emitters, such as $^{140}$Nd and $^{161}$Tb, for which the Auger energy spectra have never been published.

The main objective of the present study was to benchmark the Auger-cascade model (Lee et al. 2012) against the literature data of three selected iodine radionuclides and provide the full Auger energy spectrum of $^{124}$I decay for the first time. $^{123}$I and $^{125}$I were chosen because they were the most studied Auger-emitting radionuclides in the past four decades medically and computationally. $^{124}$I, which is the only long-lived positron-emitting radioisotope of iodine, has been shown to be a potentially good Auger-emitter by Nikjoo et al (2008). Herein, we present comparisons of the calculated average emission spectra of $^{123}$I and $^{125}$I with the previously published values as well as a comparison of the average emission spectra of $^{124}$I calculated in isolated-atom and condensed-phase approximations. A comparison of calculated Auger and Coster-Kronig multiplicities per decay with the literature values is also presented for all three radionuclides.

## Materials and Methods

A detailed description of the simulation code used for this paper has been given elsewhere (Lee et al. 2012, 2015); here only the important aspects will be summarized. Initial vacancy distributions due to EC and IC were evaluated using Schönfeld's approach (Schönfeld 1998) and BRICC (Kibédi et al. 2008), respectively, based on the nuclear structure data from the Evaluated Nuclear Structure Data File (ENSDF). In the first step, a random number was generated to determine the EC branch by which the parent nucleus decays. A subshell was selected to be ionized by the next random number according to the initial vacancy distribution calculated for the chosen EC branch. Energies of all possible radiative and non-radiative transitions to fill the chosen vacancy were checked and any energetically forbidden transition were discarded. A transition energy was calculated as the difference between the total energies of the atom before and after an event using the RAINE code (Band et al. 2002), which is based on the relativistic Dirac-Fock (DF) method. Transition probabilities were obtained from the Evaluated Atomic Data Library (EADL) (Perkins et al. 1991). Since all atomic transition probabilities were calculated for singly ionized atoms, an empirical correction, which was first used by Krause and Carlson (1967), was applied to modify them according to the actual number of electrons available in the participating subshells. The transition probabilities of KLL Auger were corrected based on the more accurate data presented by Chen et al. (1980). Next, a random number was generated to determine the transition to fill the chosen vacancy out of the list of energetically allowed transitions.

The protocol was similar for scenarios when there are multiple vacancies. Energies of all transitions from EADL were checked for each vacancy before a transition was randomly selected from the list of remaining energetically allowed transitions. For reasons to be explained below, the simulation of the Auger cascade was terminated when there were no more energetically allowed transition, or transitions associated with lifetimes larger than $10^{-12}$ s.

After the first Auger cascade, a nuclear transition was selected, based on the next random number, to proceed from the excited level that the previously chosen EC branch had decayed to. If an IC process was selected, an atomic subshell was randomly chosen to be ionized based on the initial vacancy distribution due to the selected IC process. A new Auger cascade was initiated after the ionization. The simulation of the nuclear decay was concluded when the daughter nucleus had reached its ground state through a cascade of nuclear transitions.

The atom was assumed to recover its neutral atomic configuration in between two successive nuclear transitions in the model. This assumption is not valid in rare cases wherein the level half-life of the daughter nucleus (~ $10^{-14}$) is comparable to the lifetime of the Auger cascade. An example of one of these rare cases was reported by Bulgakov et al. (1987).

Two different approaches toward the fate of valence vacancies *during* the Auger cascade have been considered in the literature, namely the isolated atom and condensed phase conditions. The distinction between the two approaches is the assumption on the neutralization of any vacancy created in the valence shell *during* an Auger cascade. In the condensed phase approach, charge transfer between the environment and the valence shell is allowed to take place during the Auger cascade, leaving the atom completely neutralized at the end of the cascade process. There is no consensus in the literature on which approach should be adopted, due to the lack of experimental evidence for the time-scale of electron transfer from DNA or proteins to an Auger-emitter. This is one of the main contributing factors leading to the large scatter in the published emission spectra of selected Auger-emitters. In this work, no preference was made and emission spectra of both approaches were simulated for detailed comparison.

The relationship between the atomic radiation yields and the maximum allowed mean lifetime of a vacancy for the isolated $^{125}I$ is illustrated in Figure 1. In this figure, mean vacancy lifetimes for singly ionized atoms were extracted from EADL, with the Krause and Carlson correction applied where necessary. The figure shows that all atomic transitions except low-energy N-shell X-rays, are released when the maximum allowed mean lifetime is set to $10^{-12}$ s. In addition, some medical radionuclides have intermediate nuclear states with lifetimes in the order of $10^{-11}$ s, for example $^{135}La$. Therefore, it was assumed in this work that any vacancy with a mean lifetime of more than $10^{-12}$ s would not be filled by atomic transitions (i.e. the Auger cascade would be terminated when all remaining vacancies have mean lifetime longer than $10^{-12}$ s). Such a vacancy was assumed to be neutralized by the environment before the subsequent nuclear transition took place.

The role of shake-off electrons in the Auger cascade and the fact that the inclusion of shake-off transitions was needed to produce the measured yields of highly charged ions in Xenon following the photoionization of the $M_{4,5}$ subshells were acknowledged in Viefhaus et al. (2005). In addition, shake-off electrons made up of about 12 % of the emitted electrons in the decays of $^{99m}Tc$ and $^{123}I$, according to the calculations performed by Pomplun (2012). However, shake-off transitions were ignored in our work since the shake-off probabilities are only available for noble gases. It is not possible to extrapolate the literature values to elements other than noble gases.

## Results

The decay spectra of $^{123}I$, $^{124}I$ and $^{125}I$ were calculated using the methodology described above with $10^5$ Monte Carlo simulations. $^{123}I$ decays with a half-life of 13.2 hours by EC (100%) followed by $\gamma$ emission (84%) and IC (16%) to $^{123}Te$. Similarly, $^{125}I$ decays with a half-life of 60.1 days by EC followed by $\gamma$ emission (7%) and IC (93%) to $^{125}Te$. $^{124}I$ decays with a half-life of 4.18 days via either positron emission (22.7%) or EC (77.3%) to excited states (65%) or the ground state (35%) of $^{124}Te$. The excited states of $^{124}Te$ de-excite in many steps to the ground state mostly by $\gamma$ emission. A small number of conversion electrons (0.35%) is also generated during this de-excitation.

### *Initial vacancy distribution*

Tables I and II compare the initial vacancy distributions calculated for three iodine isotopes to values in the literature. Contributions from subshells other than the first two subshells in each principal atomic shell were neglected in this work as Schönfeld's approach (1998) works only for allowed and non-unique first-forbidden EC transitions and these transitions have negligible capture probabilities to the aforementioned subshells. Other types of EC decay are uncommon and weak in strength thus they were ignored in this work, except for unique first forbidden transitions. This type of transition was assumed to be the same as the non-unique first-forbidden transition. Among the three iodine radionuclides, only $^{124}I$ has non-negligible unique first forbidden transitions and the contribution from $L_3$ subshell in $^{124}I$ is 1.7% according to Nikjoo et al. (2008). Humm (1984) and Pomplun (1992) included the contributions from $L_3$ and $M_3$ subshells in $^{123}I$ even though only allowed and first non-unique forbidden EC transitions are present in the radionuclide. The calculated initial vacancy distributions due to IC processes are consistent with the literature values.

## *Average spectra*

### $^{123}$*I and* $^{125}$*I*

Tables III and V display the calculated average decay spectra of $^{123}$I and $^{125}$I for the isolated atom and condensed phase approximations. Nuclear radiations with yields of less than $10^{-3}$ have been excluded. The results of Pomplun (2012), Howell (1992), and Stepanek (2000) are included for comparison.

Pomplun's method was almost equivalent to ours except that he included shake-off transitions by extrapolating shake-off probabilities of noble gases. In his calculations, some outer-shell electrons were released as shake-off transitions, thus reducing the number of Auger NXY significantly. However, both methods agree on the total number of electrons released per decay (see Table VI below).

Transition energies reported by Stepanek were calculated in a similar way to our method. However, his N-shell Auger yields in the decay of $^{125}$I are significantly lower. In his calculations, the atom was assumed not to recover its neutral atomic configuration after the first Auger cascade, thus the atom remained highly ionized when the second Auger cascade took place. In the case of $^{125}$I, the excited state of the daughter nucleus $^{125}$Te has a half-life of 1.48 ns, which means that no interaction between the atom and the environment for at least $10^{-9}$ s was assumed in Stepanek's model.

For $^{123}$I and $^{125}$I in the condensed-phase approximations, there is a good agreement in terms of radiation yields between the present spectra and those of Howell up to N-shell transitions. Calculated average energies of outer-shell transitions in this work are consistently lower than Howell's values as the $(Z + 1)/Z$ rule (Chung and Jenkins 1970) was adopted in his calculations. The $(Z + 1)/Z$ rule does not account for the existence of multiple vacancies and thus it could overestimate the transition energies when there are multiple vacancies. Howell included O-shell non-radiative transitions in his calculations, resulting in an extra 2.18 and 3.66 very low-energy OOX transitions on average for $^{123}$I and $^{125}$I, respectively. These transitions are not present in EADL (Perkins et al. 1991) and it is debatable to include them in the simulations. O-shell non-radiative transitions of singly ionized xenon, which has two more electrons than singly ionized tellurium, were determined to be energetically forbidden in the relativistic Dirac-Fock method by using computer codes GRASP2K (Jönsson et al. 2013) and RATIP (Fritzsche 2012). This indicates that the OOX transitions in tellurium are unphysical, therefore they were not considered in this work.

*$^{124}$I*

Table IV shows the average spectrum of $^{124}$I for the isolated-atom and condensed-phase approximations. Nuclear radiations with yields of less than $10^{-2}$ have been excluded. The two approximations deviate from each other only for N-shell transitions. In general, the condensed phase leads to an enhancement in the emission of N-shell Auger and Coster-Kronig transitions. In comparison, some N-shell non-radiative transitions become energetically forbidden in the isolated atom due to the build up of charge during the Auger cascade. This leads to greater yields of N-shell X-rays as they remain energetically possible near the end of the Auger cascade.

***Examples of Monte Carlo calculated spectra of radiations released in the decays of $^{123}$I, $^{124}$I and $^{125}$I***

In Appendices A, B and C are given 15 examples of radiations released in independent decays of $^{123}$I, $^{124}$I and $^{125}$I, respectively, in the condensed phase. Column 1 indicates the decay number and columns 2 and 3 give the type and the number of radiations released in that decay. Columns 4 to 11 give the transition energies in eV. These examples show that every independent decay is different and large variations are observed in the energy and the number of electrons released in each decay.

*Kinetic energy spectra & Frequency distributions*

Figure 2 shows the kinetic energy spectra of photons and electrons for the decays of $^{123}$I, $^{124}$I and $^{125}$I in the isolated atom and condensed phase. For the decay of $^{125}$I, a total of 99742 and 180942 distinct electron energies were recorded from $10^5$ simulations for the isolated atom and condensed phase, respectively. Different kinetic energies can be carried away by the same atomic transition due to the stochastic nature of the Auger cascade; the exact amount of energy is determined by the difference in the total energies of the atom before and after the transition, which are determined by the number and the location of co-existing vacancies. No difference between the two conditions is observed for energies above 1 keV. However, there are large discrepancies at energies below 100 eV. This is the region of N-shell transitions, which strongly depend on the electron population of the valence shell. This observation is consistent with the results published by Pomplun (2000).

Figure 3 displays the frequency distributions of the number of emitted electrons and photons for $^{123}$I, $^{124}$I and $^{125}$I in the isolated atom and condensed phase. Figure 3a-3c show that many more electrons can be emitted per decay of radioiodine when the atoms are allowed to have an extra supply of electrons from the environment during the Auger cascade in the condensed phase. In comparison, there is no difference between the two approximations for the frequency distributions of the emitted photons. Most of the low-energy radiative transitions have a long mean lifetime ($\geq 10^{-7}$ s) thus they were excluded in this work. If these transitions were taken into account, the isolated atom would be expected to yield more photons thus widening the photon frequency distributions.

A zigzag pattern is observed in the frequency distributions of the number of emitted electrons in the condensed phase (see Figure 3a-c). This observation is a consequence of the condensed-phase approximation and the fact that non-radiative transitions dominate in the atomic shells other than K-shell. The condensed-phase approximation can prevent the non-radiative transitions from becoming energetically forbidden through the neutralization of valence vacancies. It is likely to have only non-radiative transitions after a non K-shell vacancy is created. Examples 4 and 12 in Appendix A show that only non-radiative transitions are present after a non K-shell initial vacancy is created due to the EC decay of $^{123}$I. The number of vacancies is multiplied by two after filling out all the vacancies created in the previous propagation step by non-radiative transitions thus increasing the number of emitted electrons by an even number. Therefore, the Auger cascade favours odd number of emitted electrons in the condensed-phase approximation. Examples 1, 5, 6, 7, 8, 10, 11, 13 in Appendix A show that the Auger cascade following EC predominantly emits odd number of electrons after the emission of a K-shell X-ray. In contrast, the Auger cascade initiated by IC prefers even number of emitted electrons after the inclusion of conversion electrons. The comparison of the frequency distributions of the number of emitted electrons calculated in the condensed-phase approximation following the EC and IC for isotopes of iodine, bromine, antimony and barium is shown in Figure 4a-d. Only the frequency distributions for the isotopes of barium do not exhibit similar zigzag pattern. This is due to the fact that barium and caesium only have 2 and 1 P-shell valence electrons, respectively, in the condensed-phase approximation. The neutralization of valence vacancies in the two elements would not be sufficient to prevent some non-radiative transitions from becoming energetically forbidden.

## Discussion

Research concerned with targeted Auger therapy requires the emission spectra of Auger-electron emitting radionuclides to evaluate dose distributions around the Auger emitting radiopharmaceuticals. To date, most researchers rely on information available in the Medical Internal Radiation Dosimetry (MIRD) monograph (Eckerman & Endo 2007) as the database for emission spectra while the computer codes used to simulate the decay of Auger emitting radionuclides are not accessible to the users (Nikjoo et al. 2008).

This paper describes a computational model which can be used to simulate the decay of Auger emitting radionuclides based on the latest nuclear structure data in the format of ENSDF. The model can simulate the emission spectra for not only the common, but also the emerging radionuclides in which only the atomic transitions from the K and L-shells were reported in MIRD. The emission specta of $^{123}$I, $^{124}$I and $^{125}$I were calculated and compared with the published results from several authors. In general, good agreement with the literature was achieved except for the outer-shell Auger and Coster-Kronig transitions.

The question on whether charge neutralization happens during the Auger cascade ($10^{-16}$ - $10^{-12}$ s) has been one of the key concerns in the modelling of Auger cascade, and the effect of this so-called fast neutralization was the only difference between the isolated atom and condensed phase approximations in this work. No preference was made in this work due to the lack of direct experimental evidence to indicate which approximation is better. Based on simulations performed in this work, major differences in the emission spectra for the two approximations are found at energies below 100 eV. Future dosimetry studies will be needed to quantify the dose enhancement due to the extra electrons released in the condensed phase.

Figure1 shows that all non-radiative transitions are completed before $10^{-12}$ s while most of the N-shell radiative transitions require $10^{-7}$ s to finish within the isolated atom approximation. If the condensed phase is valid, the atom would require an electron transfer rate of at least $10^{12}$ $s^{-1}$ from the environment to an Auger emitter. Page *et al* (Page et al. 1999) predicted that the electron transfer rate within oxidoreductase proteins is above $10^{12}$ $s^{-1}$ when the distance between two atoms, which are covalently bonded, is less than 5 Å. This suggests that the validity of the condensed phase approximation could depend on the structure of the radiopharmaceutical that incorporates the Auger-electron emitter.

A comparison of simulated Auger and Coster-Kronig yields with the literature values for the three iodine radionuclides is shown in Table VI. Excellent agreement with results from Nikjoo et al. (2008) is obtained even though two very different methods of calculating transition energies were used. This agreement shows that the $(Z+1)/Z$ rule, which was used by Nikjoo et al. as well as Howell (1992), is a good approximation for Auger electron energies in condensed phase condition, even though it does not account for a realistic vacancy distribution. Due to the presence of multiple vacancies during the Auger cascade, one Auger electron transition can carry away many different kinetic energies, but this is ignored by the $(Z+1)/Z$ rule. The changes in transition energies need to be carefully determined and accounted for since the interaction of low-energy electrons with water (≤ 1 keV) is highly dependent of their incident energy (Nikjoo et al. 2008). Howell (1992) agreed that each Auger electron energy should be calculated based on the electron configuration for that particular event but he chose the $(Z+1)/Z$ rule due to the extensive CPU time required for the strict energy bookkeeping. With such extensive bookkeeping, the simulated L-Auger energy spectrum of $^{131}$Cs using our model was shown to have near-perfect agreement with experiment (Lee et al. 2013).

To date, the EADL has been the main input of Auger-cascade simulations, but its validity is less well-established. Recent systematic and quantitative validation at CERN indicated that the K- and L- shell X-ray transition probabilities from EADL do not reflect the state-of-the-art (Pia et al. 2009). In an earlier study performed by Stepanek (2000), Auger and Coster-Kronig transitions from EADL were shown to have poor agreement with experiments for elements with $Z < 60$. Furthermore, the 2011 IAEA special meeting on Intermediate-term Nuclear Data Needs for Medical Applications (Nichols et al. 2011) reported: "*A comprehensive calculational route also needs to be developed to determine the energies and emission probabilities of the low-energy X-rays and Auger electrons to a higher degree of detail and consistency than is available at present.*" These reports strongly suggested that some contents of EADL need to be updated. One way is to replace atomic data from EADL with the relativistic Dirac-Fock (DF) calculations using the state-of-the-art computer codes GRASP2K (Jönsson et al. 2013) and RATIP (Fritzsche 2012). A set of systematic DF calculations could determine whether the missing transitions in EADL should be included in the Auger-cascade simulations, for example, the missing OOX transitions in tellurium have been determined to be energetically forbidden.  Preliminary results using the DF transition probabilities indicated non-trivial differences in energies and yields of emitted atomic radiations as compared to EADL. Work is in progress to quantify the impact of these difference.

Although good agreement with the literature calculations was achieved, the model needs to be benchmarked against experimental data in the near future. One way to validate the model is to systematically compare simulated residual ionic charge distributions with experiments. Ionic charge-state measurements at the LOHENGRIN fission-fragment separator could provide valuable experimental data for unambiguous benchmarking (Wohlfarth et al. 1978). Collaborative work with researchers at the LOHENGRIN separator, to validate the model, is in progress.

## Conclusions

A recently developed Monte Carlo model of the Auger cascade was benchmarked against the literature for the decays of $^{123}I$, $^{124}I$ and $^{125}I$. Good agreement was achieved with previously published values, except for the outer-shell Auger and Coster-Kronig transitions. Nevertheless, further benchmarking of the model against experimental data is needed to fine-tune the model.

## Acknowledgements

This work was supported by the Australian Research Council Discovery grant (DP140103317).

## Declaration of interest

The authors report no conflict of interest. The authors alone are responsible for the content and writing of the paper.

**Figures**

Figure 1.

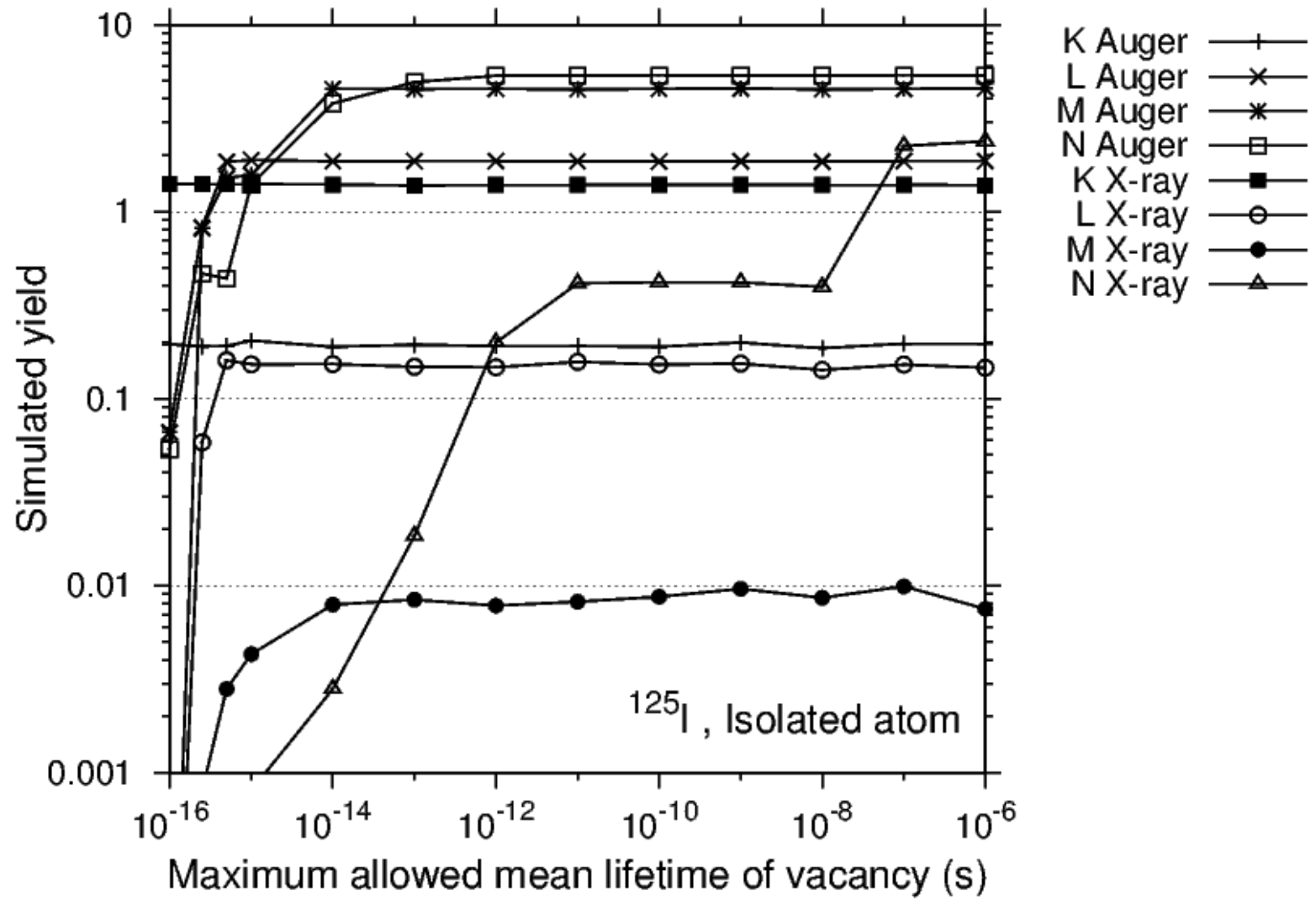

Figure 1: Simulated atomic radiation yields following the decay of $^{125}I$ as a function of the maximum allowed mean lifetime of vacancy in the isolated-atom approach. Given that most of the atomic radiations are released when the maximum allowed mean lifetime is set to $10^{-12}$ s and some medical radionuclides have short-lived intermediate nuclear states (~ $10^{-11}$ s), it was assumed that any vacancy with mean lifetime longer than $10^{-12}$ s would be ignored during the cascade and such a vacancy was assumed to be neutralized by the environment (see text for details). Coster-Kronig transitions are included in the labelled Auger transitions.

Figure 2.

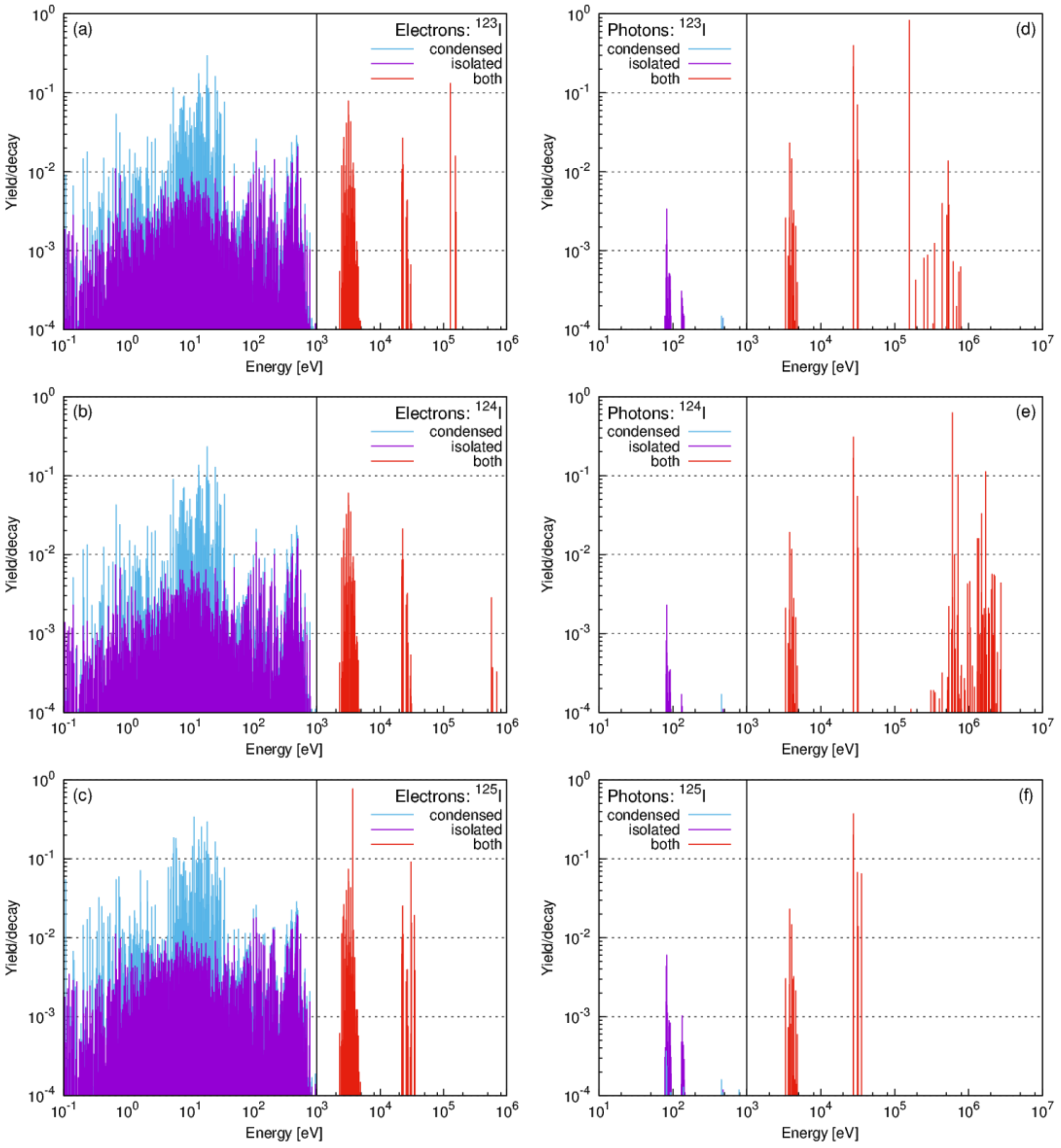

Figure 2: Energy spectra of photons and electrons of $^{123}$I (a, d), $^{124}$I (b, e) and $^{125}$I (c, f) calculated in the isolated-atom (light blue) and condensed-phase (purple) approximations. Since the two approximations have negligible differences for energies above 1 keV, results from the isolated-atom are used to present both approximations in this region (red). At energies below 1000 eV, electron spectra for the condensed phase are plotted behind their counterparts while photon spectra for the same approximation are plotted in front of their counterparts. These spectra show that there are large discrepancies at energies below 100 eV.

Figure 3.

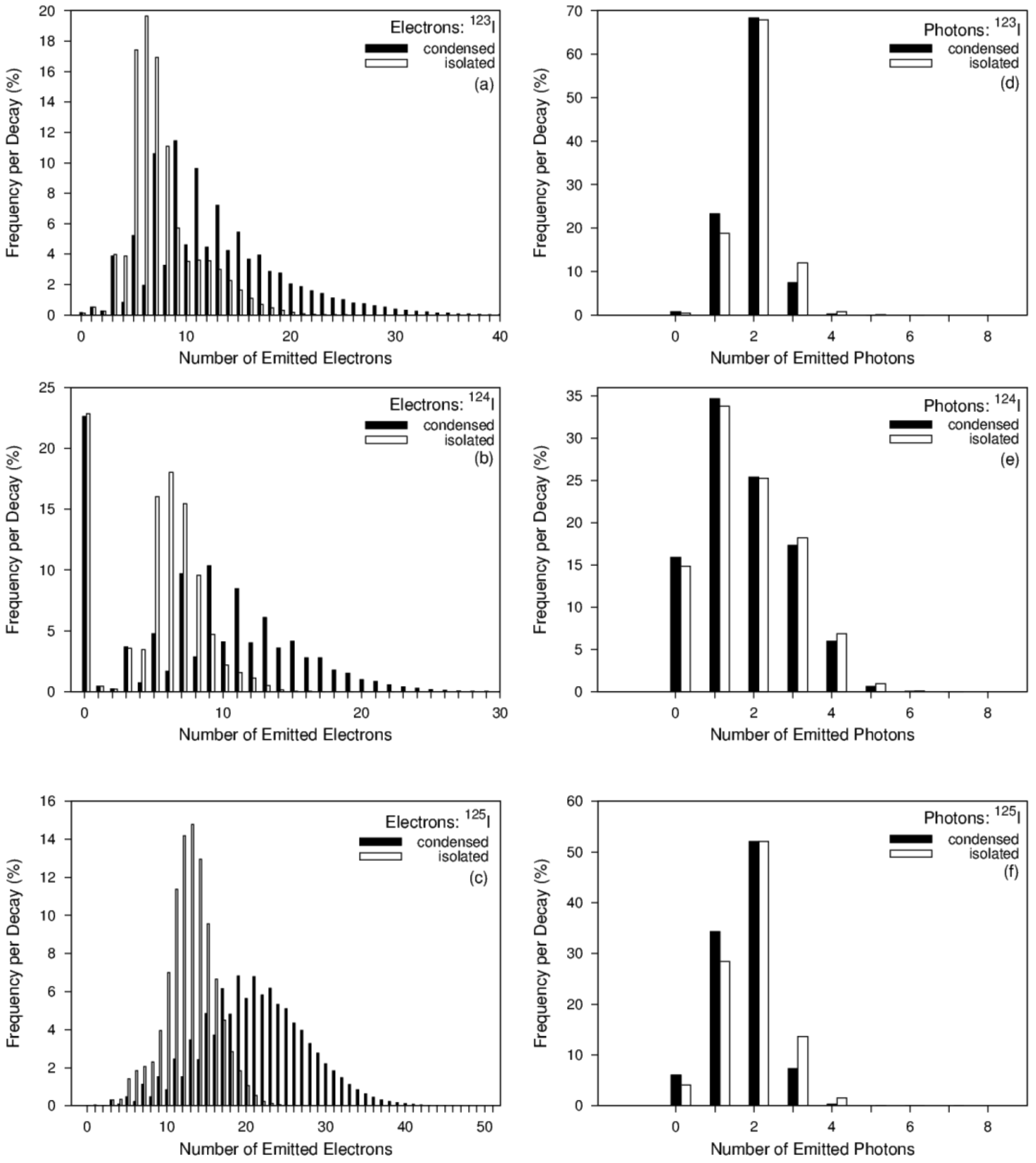

Figure 3: Frequency distributions of the number of emitted electrons and photons for $^{123}$I (a, d), $^{124}$I (b, e) and $^{125}$I (c,f) in the isolated-atom and condensed-phase approximations. The condensed phase yields wider distributions than the isolated atom for electrons while the two approximations give very similar distributions for photons. A huge spike at 0 in figure 3b is due to the fact that $^{124}$I is also a positron emitter (22.7%).

Figure 4.

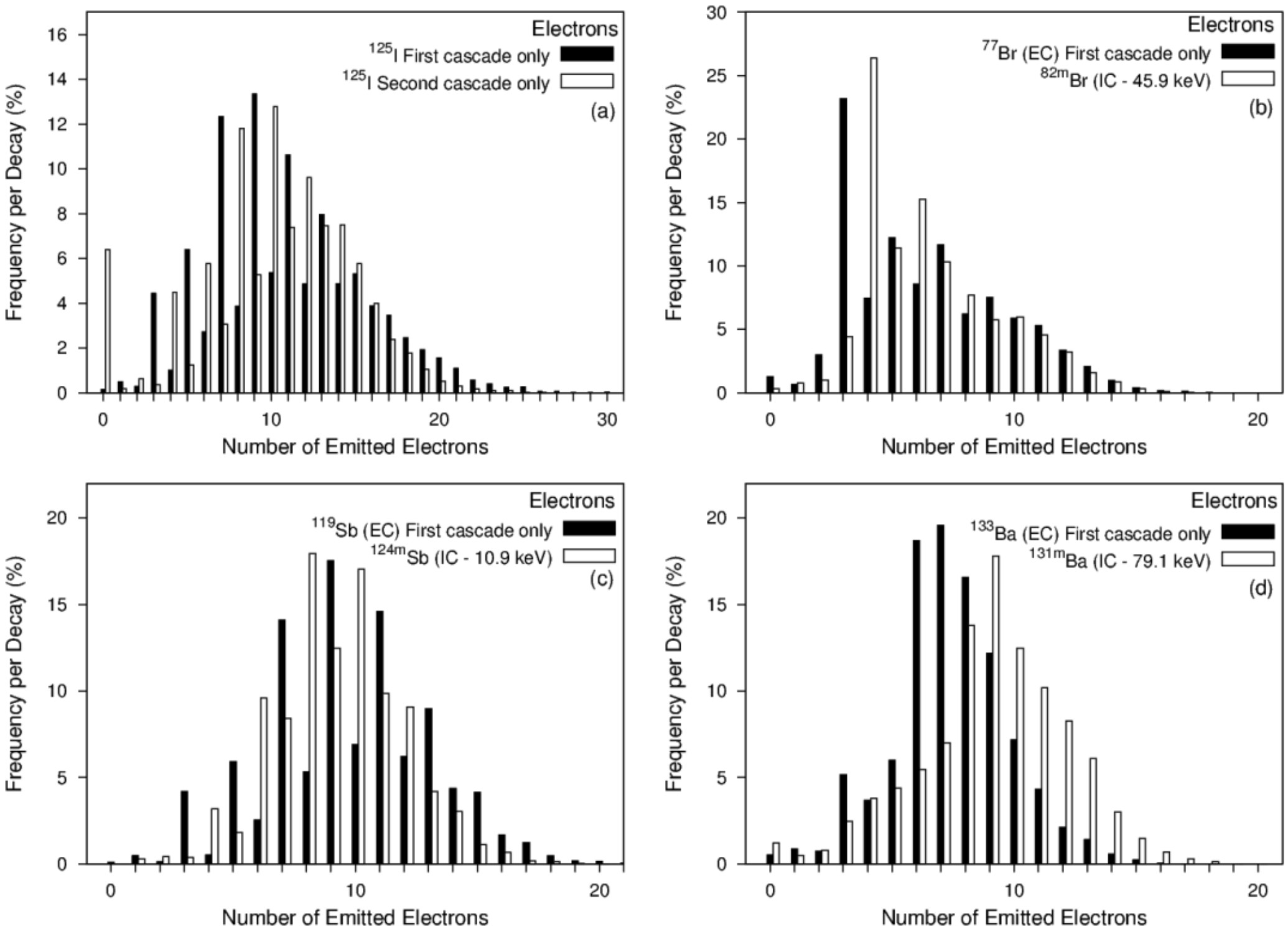

Figure 4: Frequency distributions of the number of emitted electrons for isotopes of iodine (a), bromine (b), antimony (c) and barium (d) in the condensed-phase approximations. Figures a-c show that the Auger cascade following EC favours odd number of emitted electrons while the Auger cascade following IC prefers even number of emitted electrons. Unlike the other elements, the frequency distributions of barium isotopes do not exhibit a zigzag pattern due to the lack of valence electrons.

## Tables

Table I: Initial vacancy distributions for $^{123}I$.

| | Pomplun et al. (1992) | | Humm (1984) | | This work | |
|---|---|---|---|---|---|---|
| **Atomic Subshell** | EC | IC | EC | IC | EC | IC |
| **K** | 0.797 | 0.135 | 0.818 | 0.147 | 0.854 | 0.134 |
| **L1** | 0.104 | 0.016 | 0.106 | 0.017 | 0.1127 | 0.016 |
| **L2** | 0.003 | 0.0011 | 0.003 | 0.002 | 0.0029 | 0.0011 |
| **L3** | 0.007 | 0.0004 | 0.007 | 0.001 | – | 0.0003 |
| **M1** | 0.022 | 0.0035 | 0.022 | 0.003 | 0.024 | 0.003 |
| **M2** | 0.001 | 0.00 | 0.001 | 0.00 | 0.0007 | 0.0002 |
| **M3** | 0.038 | 0.00 | 0.038 | 0.00 | – | 0.00 |
| **M4** | 0.00 | 0.00 | 0.00 | 0.00 | – | 0.00 |
| **M5** | 0.00 | 0.00 | 0.00 | 0.00 | – | 0.00 |
| **N1** | 0.005 | 0.0008 | 0.005 | 0.00 | 0.0052 | 0.0006 |
| **N2** | 0.00 | – | – | – | 0.0002 | 0.00 |
| **N3** | 0.00 | – | – | – | – | 0.00 |
| **O1** | – | – | – | – | 0.0003 | 0.00 |
| **Sum** | 0.977 | 0.1568 | 1.0 | 0.17 | 1.0 | 0.1552 |

Table II: Initial vacancy distributions for $^{124}I$ and $^{125}I$.

| **Radionuclide** | $^{124}I$ | | | $^{125}I$ | | | |
|---|---|---|---|---|---|---|---|
| | Nikjoo (2008) | This work | | Charlton and Booz (Nikjoo et al. 2008) | | This work | |
| **Atomic Subshell** | EC | EC | IC | EC | IC | EC | IC |
| **K** | 0.644 | 0.661 | 0.00303 | 0.806 | 0.795 | 0.8024 | 0.776 |
| **L1** | 0.105 | 0.0867 | 0.00034 | 0.154 | 0.095 | 0.1512 | 0.0936 |
| **L2** | 0.003 | 0.0023 | 0.00004 | 0.004 | 0.007 | 0.0039 | 0.0172 |
| **L3** | 0.017 | – | 0.00003 | – | 0.004 | – | 0.0158 |
| **M1** | 0.004 | 0.0184 | 0.00007 | 0.0035 | 0.019 | 0.0336 | 0.0186 |
| **M2** | – | 0.0005 | 0.00 | 0.001 | 0.004 | 0.0009 | 0.0036 |
| **M3** | – | – | 0.00 | – | 0.001 | – | 0.0033 |
| **M4** | – | – | 0.00 | – | – | – | 0.00 |
| **M5** | – | – | 0.00 | – | – | – | 0.00 |
| **N1** | – | 0.004 | 0.00001 | – | 0.004 | 0.0074 | 0.0037 |
| **N2** | – | 0.0001 | 0.00 | – | 0.001 | 0.0002 | 0.0007 |
| **N3** | – | – | 0.00 | – | – | – | 0.0006 |
| **O1** | – | 0.0002 | 0.00 | – | – | 0.0004 | 0.0004 |
| **Sum** | 0.773 | 0.7732 | 0.00352 | 1.0 | 0.930 | 1.0 | 0.9335 |

Table III: Average spectrum per decay for isolated atom and condensed phase of $^{123}$I.

| | **Isolated atom** | | | | **Condensed phase** | | | |
|---|---|---|---|---|---|---|---|---|
| | This work | | Pomplun (2012) | | This work | | Howell (1992) | |
| **Process** | Average energy (eV) | Yield | Average energy (eV) | Yield | Average energy (eV) | Yield | Average energy (eV) | Yield |
| $\gamma$**1** | $1.59 \cdot 10^{5}$ | $8.36 \cdot 10^{-1}$ | $1.59 \cdot 10^{5}$ | $8.45 \cdot 10^{-1}$ | $1.59 \cdot 10^{5}$ | $8.38 \cdot 10^{-1}$ | $1.59 \cdot 10^{5}$ | $8.39 \cdot 10^{-1}$ |
| **IC1 K** | $1.27 \cdot 10^{5}$ | $1.34 \cdot 10^{-1}$ | $1.27 \cdot 10^{5}$ | $1.28 \cdot 10^{-1}$ | $1.27 \cdot 10^{5}$ | $1.33 \cdot 10^{-1}$ | $1.27 \cdot 10^{5}$ | $1.30 \cdot 10^{-1}$ |
| **IC1 L** | $1.54 \cdot 10^{5}$ | $1.73 \cdot 10^{-2}$ | $1.54 \cdot 10^{5}$ | $1.71 \cdot 10^{-2}$ | $1.54 \cdot 10^{5}$ | $1.71 \cdot 10^{-2}$ | $1.54 \cdot 10^{5}$ | $1.79 \cdot 10^{-2}$ |
| **IC1 M+** | $1.58 \cdot 10^{5}$ | $4.11 \cdot 10^{-3}$ | $1.58 \cdot 10^{5}$ | $3.80 \cdot 10^{-3}$ | $1.58 \cdot 10^{5}$ | $4.42 \cdot 10^{-3}$ | $1.58 \cdot 10^{5}$ | $5.30 \cdot 10^{-3}$ |
| $\gamma$**3** | $4.40 \cdot 10^{5}$ | $4.00 \cdot 10^{-3}$ | $4.40 \cdot 10^{5}$ | $3.00 \cdot 10^{-3}$ | $4.40 \cdot 10^{5}$ | $4.03 \cdot 10^{-3}$ | $4.40 \cdot 10^{5}$ | $3.90 \cdot 10^{-3}$ |
| $\gamma$**5** | $3.46 \cdot 10^{5}$ | $1.25 \cdot 10^{-3}$ | $3.46 \cdot 10^{5}$ | $7.00 \cdot 10^{-4}$ | $3.46 \cdot 10^{5}$ | $1.06 \cdot 10^{-3}$ | $3.46 \cdot 10^{5}$ | $1.40 \cdot 10^{-3}$ |
| $\gamma$**6** | $5.05 \cdot 10^{5}$ | $2.85 \cdot 10^{-3}$ | $5.05 \cdot 10^{5}$ | $2.50 \cdot 10^{-3}$ | $5.05 \cdot 10^{5}$ | $2.87 \cdot 10^{-3}$ | $5.05 \cdot 10^{5}$ | $2.80 \cdot 10^{-3}$ |
| $\gamma$**12** | $5.29 \cdot 10^{5}$ | $1.38 \cdot 10^{-2}$ | $5.29 \cdot 10^{5}$ | $1.13 \cdot 10^{-2}$ | $5.29 \cdot 10^{5}$ | $1.38 \cdot 10^{-2}$ | $5.29 \cdot 10^{5}$ | $1.45 \cdot 10^{-2}$ |
| $\gamma$**17** | $5.39 \cdot 10^{5}$ | $3.85 \cdot 10^{-3}$ | $5.39 \cdot 10^{5}$ | $2.80 \cdot 10^{-3}$ | $5.39 \cdot 10^{5}$ | $3.82 \cdot 10^{-3}$ | $5.39 \cdot 10^{5}$ | $4.50 \cdot 10^{-3}$ |
| **Auger KLL** | $2.26 \cdot 10^{4}$ | $8.04 \cdot 10^{-2}$ | $2.25 \cdot 10^{4}$ | $7.31 \cdot 10^{-2}$ | $2.26 \cdot 10^{4}$ | $8.14 \cdot 10^{-2}$ | $2.24 \cdot 10^{4}$ | $8.38 \cdot 10^{-2}$ |
| **Auger KLX** | $2.65 \cdot 10^{4}$ | $3.49 \cdot 10^{-2}$ | $2.64 \cdot 10^{4}$ | $3.28 \cdot 10^{-2}$ | $2.65 \cdot 10^{4}$ | $3.57 \cdot 10^{-2}$ | $2.63 \cdot 10^{4}$ | $3.84 \cdot 10^{-2}$ |
| **Auger KXY** | $3.04 \cdot 10^{4}$ | $3.73 \cdot 10^{-3}$ | $3.03 \cdot 10^{4}$ | $2.80 \cdot 10^{-3}$ | $3.04 \cdot 10^{4}$ | $4.15 \cdot 10^{-3}$ | $3.02 \cdot 10^{4}$ | $3.50 \cdot 10^{-3}$ |
| **CK LLX** | $2.84 \cdot 10^{2}$ | $1.53 \cdot 10^{-1}$ | $2.80 \cdot 10^{2}$ | $1.34 \cdot 10^{-1}$ | $2.88 \cdot 10^{2}$ | $1.55 \cdot 10^{-1}$ | $2.13 \cdot 10^{2}$ | $1.56 \cdot 10^{-1}$ |
| **Auger LMM** | $3.05 \cdot 10^{3}$ | $7.34 \cdot 10^{-1}$ | $3.03 \cdot 10^{3}$ | $7.11 \cdot 10^{-1}$ | $3.05 \cdot 10^{3}$ | $7.32 \cdot 10^{-1}$ | $3.04 \cdot 10^{3}$ | $7.51 \cdot 10^{-1}$ |
| **Auger LMX** | $3.68 \cdot 10^{3}$ | $2.08 \cdot 10^{-1}$ | $3.66 \cdot 10^{3}$ | $2.00 \cdot 10^{-1}$ | $3.68 \cdot 10^{3}$ | $2.08 \cdot 10^{-1}$ | $3.66 \cdot 10^{3}$ | $2.02 \cdot 10^{-1}$ |
| **Auger LXY** | $4.30 \cdot 10^{3}$ | $1.45 \cdot 10^{-2}$ | $4.31 \cdot 10^{3}$ | $1.22 \cdot 10^{-2}$ | $4.32 \cdot 10^{3}$ | $1.50 \cdot 10^{-2}$ | $4.28 \cdot 10^{3}$ | $1.30 \cdot 10^{-2}$ |
| **CK MMX** | $9.67 \cdot 10^{1}$ | $7.77 \cdot 10^{-1}$ | $9.24 \cdot 10^{1}$ | $8.70 \cdot 10^{-1}$ | $1.03 \cdot 10^{2}$ | $7.91 \cdot 10^{-1}$ | $1.27 \cdot 10^{2}$ | $8.69 \cdot 10^{-1}$ |
| **Auger MXY** | $4.11 \cdot 10^{2}$ | $1.94 \cdot 10^{+0}$ | $3.94 \cdot 10^{2}$ | $1.93 \cdot 10^{+0}$ | $4.27 \cdot 10^{2}$ | $1.94 \cdot 10^{+0}$ | $4.61 \cdot 10^{2}$ | $1.97 \cdot 10^{+0}$ |
| **CK NNX** | $2.01 \cdot 10^{1}$ | $1.35 \cdot 10^{+0}$ | $2.85 \cdot 10^{1}$ | $1.27 \cdot 10^{+0}$ | $2.19 \cdot 10^{1}$ | $2.00 \cdot 10^{+0}$ | $2.98 \cdot 10^{1}$ | $2.10 \cdot 10^{+0}$ |
| **Auger NXY** | $2.08 \cdot 10^{1}$ | $2.10 \cdot 10^{+0}$ | $3.02 \cdot 10^{1}$ | $1.13 \cdot 10^{+0}$ | $1.49 \cdot 10^{1}$ | $6.36 \cdot 10^{+0}$ | $3.25 \cdot 10^{1}$ | $6.54 \cdot 10^{+0}$ |
| **CK OOX** | – | – | – | – | – | – | $6.00 \cdot 10^{0}$ | $2.18 \cdot 10^{+0}$ |
| **X-ray K$_{\alpha}$1** | $2.76 \cdot 10^{4}$ | $4.66 \cdot 10^{-1}$ | $2.75 \cdot 10^{4}$ | $4.47 \cdot 10^{-1}$ | $2.76 \cdot 10^{4}$ | $4.63 \cdot 10^{-1}$ | $2.75 \cdot 10^{4}$ | $4.62 \cdot 10^{-1}$ |
| **X-ray $K_{\alpha}2$** | $2.73 \cdot 10^{4}$ | $2.51 \cdot 10^{-1}$ | $2.72 \cdot 10^{4}$ | $2.48 \cdot 10^{-1}$ | $2.73 \cdot 10^{4}$ | $2.50 \cdot 10^{-1}$ | $2.72 \cdot 10^{4}$ | $2.37 \cdot 10^{-1}$ |
| **X-ray $K_{\beta}1$** | $3.11 \cdot 10^{4}$ | $8.17 \cdot 10^{-2}$ | $3.10 \cdot 10^{4}$ | $8.19 \cdot 10^{-2}$ | $3.11 \cdot 10^{4}$ | $8.17 \cdot 10^{-2}$ | $3.10 \cdot 10^{4}$ | $8.13 \cdot 10^{-2}$ |
| **X-ray $K_{\beta}2$** | $3.18 \cdot 10^{4}$ | $2.61 \cdot 10^{-2}$ | $3.17 \cdot 10^{4}$ | $2.24 \cdot 10^{-2}$ | $3.18 \cdot 10^{4}$ | $2.58 \cdot 10^{-2}$ | $3.17 \cdot 10^{4}$ | $2.27 \cdot 10^{-2}$ |
| **X-ray $K_{\beta}3$** | $3.10 \cdot 10^{4}$ | $4.34 \cdot 10^{-2}$ | $3.09 \cdot 10^{4}$ | $3.97 \cdot 10^{-2}$ | $3.10 \cdot 10^{4}$ | $4.36 \cdot 10^{-2}$ | $3.09 \cdot 10^{4}$ | $4.45 \cdot 10^{-2}$ |
| **X-ray KM+** | $3.17 \cdot 10^{4}$ | $2.32 \cdot 10^{-3}$ | $3.12 \cdot 10^{4}$ | $8.00 \cdot 10^{-4}$ | $3.17 \cdot 10^{4}$ | $2.35 \cdot 10^{-3}$ | $3.17 \cdot 10^{4}$ | $1.30 \cdot 10^{-3}$ |
| **X-ray L** | $3.93 \cdot 10^{3}$ | $8.90 \cdot 10^{-2}$ | $3.94 \cdot 10^{3}$ | $8.22 \cdot 10^{-2}$ | $3.93 \cdot 10^{3}$ | $8.96 \cdot 10^{-2}$ | $3.93 \cdot 10^{3}$ | $7.90 \cdot 10^{-2}$ |
| **X-ray M** | $5.41 \cdot 10^{2}$ | $4.58 \cdot 10^{-3}$ | – | – | $5.46 \cdot 10^{2}$ | $4.61 \cdot 10^{-3}$ | $5.43 \cdot 10^{2}$ | $2.30 \cdot 10^{-3}$ |
| **X-ray N** | $9.09 \cdot 10^{1}$ | $1.08 \cdot 10^{-1}$ | – | – | $1.02 \cdot 10^{2}$ | $2.81 \cdot 10^{-3}$ | – | – |

Table IV: Average spectrum per decay for isolated atom and condensed phase of $^{124}$I.

| | **Isolated atom** | | **Condensed phase** | | | |
|---|---|---|---|---|---|---|
| **Process** | Average energy (eV) | Yield | Average energy (eV) | Yield | $\Delta E/E$ (%) | $\Delta I/I$ (%) |
| $\gamma 1$ | $6.02 \cdot 10^{5}$ | $6.30 \cdot 10^{-1}$ | $6.03 \cdot 10^{5}$ | $6.31 \cdot 10^{-1}$ | 0 | +0.2 |
| $\gamma 2$ | $6.46 \cdot 10^{5}$ | $1.02 \cdot 10^{-2}$ | $6.46 \cdot 10^{5}$ | $1.00 \cdot 10^{-2}$ | 0 | -2.0 |
| $\gamma 3$ | $7.23 \cdot 10^{5}$ | $1.02 \cdot 10^{-1}$ | $7.23 \cdot 10^{5}$ | $1.03 \cdot 10^{-1}$ | 0 | +1.0 |
| $\gamma 4$ | $1.33 \cdot 10^{6}$ | $1.59 \cdot 10^{-2}$ | $1.33 \cdot 10^{6}$ | $1.60 \cdot 10^{-2}$ | 0 | +0.6 |
| $\gamma 21$ | $1.69 \cdot 10^{6}$ | $1.14 \cdot 10^{-1}$ | $1.69 \cdot 10^{5}$ | $1.13 \cdot 10^{-1}$ | 0 | -0.9 |
| $\gamma 44$ | $1.38 \cdot 10^{6}$ | $1.59 \cdot 10^{-2}$ | $1.38 \cdot 10^{5}$ | $1.61 \cdot 10^{-2}$ | 0 | +1.3 |
| $\gamma 54$ | $1.51 \cdot 10^{6}$ | $3.18 \cdot 10^{-2}$ | $1.51 \cdot 10^{5}$ | $3.33 \cdot 10^{-2}$ | 0 | +4.7 |
| **Auger KLL** | $2.26\cdot10^{4}$ | $5.41\cdot10^{-2}$ | $2.26\cdot10^{4}$ | $5.44\cdot10^{-2}$ | 0 | +0.6 |
| **Auger KLX** | $2.66\cdot10^{4}$ | $2.40\cdot10^{-2}$ | $2.66\cdot10^{4}$ | $2.43\cdot10^{-2}$ | 0 | +1.3 |
| **Auger KXY** | $3.04\cdot10^{4}$ | $2.57\cdot10^{-3}$ | $3.04\cdot10^{4}$ | $2.51\cdot10^{-3}$ | 0 | -2.3 |
| **CK LLX** | $2.85\cdot10^{2}$ | $1.03\cdot10^{-1}$ | $2.87\cdot10^{2}$ | $1.03\cdot10^{-1}$ | +0.7 | 0 |
| **Auger LMM** | $3.05\cdot10^{3}$ | $4.93\cdot10^{-1}$ | $3.05\cdot10^{3}$ | $4.94\cdot10^{-1}$ | 0 | +0.2 |
| **Auger LMX** | $3.68\cdot10^{3}$ | $1.39\cdot10^{-1}$ | $3.68\cdot10^{3}$ | $1.38\cdot10^{-1}$ | 0 | -0.7 |
| **Auger LXY** | $4.30\cdot10^{3}$ | $9.33\cdot10^{-3}$ | $4.30\cdot10^{3}$ | $9.82\cdot10^{-3}$ | 0 | +5.3 |
| **CK MMX** | $9.71\cdot10^{1}$ | $5.22\cdot10^{-1}$ | $1.04\cdot10^{2}$ | $5.33\cdot10^{-1}$ | +7.1 | +2.1 |
| **Auger MXY** | $4.13\cdot10^{2}$ | $1.30\cdot10^{+0}$ | $4.29\cdot10^{2}$ | $1.30\cdot10^{+0}$ | +3.9 | 0 |
| **CK NNX** | $1.89\cdot10^{1}$ | $9.30\cdot10^{-1}$ | $2.18\cdot10^{1}$ | $1.37\cdot10^{+0}$ | +15.3 | +47.3 |
| **Auger NXY** | $2.05\cdot10^{1}$ | $1.46\cdot10^{+0}$ | $1.51\cdot10^{1}$ | $4.35\cdot10^{+0}$ | -26.3 | +198.0 |
| **X-ray $K_{\alpha}$1** | $2.76\cdot10^{4}$ | $3.11\cdot10^{-1}$ | $2.76\cdot10^{4}$ | $3.11\cdot10^{-1}$ | 0 | 0 |
| **X-ray $K_{\alpha}$2** | $2.73\cdot10^{4}$ | $1.68\cdot10^{-1}$ | $2.73\cdot10^{4}$ | $1.68\cdot10^{-1}$ | 0 | 0 |
| **X-ray $K_{\beta}$1** | $3.11\cdot10^{4}$ | $5.57\cdot10^{-2}$ | $3.11\cdot10^{4}$ | $5.55\cdot10^{-2}$ | 0 | -0.4 |
| **X-ray $K_{\beta}$2** | $3.18\cdot10^{4}$ | $1.71\cdot10^{-2}$ | $3.18\cdot10^{4}$ | $1.80\cdot10^{-2}$ | 0 | +5.3 |
| **X-ray $K_{\beta}$3** | $3.10\cdot10^{4}$ | $2.79\cdot10^{-2}$ | $3.10\cdot10^{4}$ | $2.87\cdot10^{-2}$ | 0 | +2.9 |
| **X-ray KM+** | $3.17\cdot10^{4}$ | $1.80\cdot10^{-3}$ | $3.17\cdot10^{4}$ | $1.71\cdot10^{-3}$ | 0 | -5.0 |
| **X-ray L** | $3.94\cdot10^{3}$ | $6.00\cdot10^{-2}$ | $3.93\cdot10^{3}$ | $6.02\cdot10^{-2}$ | -0.3 | +0.3 |
| **X-ray M** | $5.64\cdot10^{2}$ | $3.39\cdot10^{-3}$ | $5.38\cdot10^{2}$ | $3.09\cdot10^{-3}$ | -4.6 | -8.8 |
| **X-ray N** | $9.07\cdot10^{1}$ | $7.17\cdot10^{-2}$ | $1.01\cdot10^{2}$ | $1.76\cdot10^{-3}$ | +11.4 | -97.5 |

Table V: Average spectrum per decay for isolated atom and condensed phase of $^{125}$I.

| | Isolated atom | | | | Condensed phase | | | |
|---|---|---|---|---|---|---|---|---|
| | This work | | Stepanek (2000) | | This work | | Howell (1992) | |
| **Process** | Average energy (eV) | Yield | Average energy (eV) | Yield | Average energy (eV) | Yield | Average energy (eV) | Yield |
| **$\gamma$1** | $3.55\cdot10^{4}$ | $6.65\cdot10^{-2}$ | $3.55\cdot10^{4}$ | $6.66\cdot10^{-2}$ | $3.55\cdot10^{4}$ | $6.52\cdot10^{-2}$ | $3.55\cdot10^{4}$ | $6.47\cdot10^{-2}$ |
| **IC1 K** | $3.68\cdot10^{3}$ | $7.75\cdot10^{-1}$ | $3.69\cdot10^{3}$ | $8.04\cdot10^{-1}$ | $3.68\cdot10^{3}$ | $7.76\cdot10^{-1}$ | $3.65\cdot10^{3}$ | $7.97\cdot10^{-1}$ |
| **IC1 L** | $3.07\cdot10^{4}$ | $1.26\cdot10^{-1}$ | $3.06\cdot10^{4}$ | $1.08\cdot10^{-1}$ | $3.07\cdot10^{4}$ | $1.28\cdot10^{-1}$ | $3.06\cdot10^{4}$ | $1.10\cdot10^{-1}$ |
| **IC1 M+** | $3.47\cdot10^{4}$ | $3.25\cdot10^{-2}$ | $3.45\cdot10^{4}$ | $2.15\cdot10^{-2}$ | $3.47\cdot10^{4}$ | $3.08\cdot10^{-2}$ | $3.47\cdot10^{4}$ | $2.84\cdot10^{-2}$ |
| **Auger KLL** | $2.26\cdot10^{4}$ | $1.28\cdot10^{-1}$ | $2.26\cdot10^{4}$ | $1.26\cdot10^{-1}$ | $2.26\cdot10^{4}$ | $1.30\cdot10^{-1}$ | $2.24\cdot10^{4}$ | $1.38\cdot10^{-1}$ |
| **Auger KLX** | $2.65\cdot10^{4}$ | $5.67\cdot10^{-2}$ | $2.65\cdot10^{4}$ | $5.80\cdot10^{-2}$ | $2.65\cdot10^{4}$ | $5.65\cdot10^{-2}$ | $2.64\cdot10^{4}$ | $5.90\cdot10^{-2}$ |
| **Auger KXY** | $3.04\cdot10^{4}$ | $5.97\cdot10^{-3}$ | $3.03\cdot10^{4}$ | $5.50\cdot10^{-3}$ | $3.04\cdot10^{4}$ | $5.96\cdot10^{-3}$ | $3.02\cdot10^{4}$ | $6.50\cdot10^{-3}$ |
| **CK LLX** | $2.87\cdot10^{2}$ | $2.69\cdot10^{-1}$ | $2.56\cdot10^{2}$ | $2.57\cdot10^{-1}$ | $2.90\cdot10^{2}$ | $2.68\cdot10^{-1}$ | $2.19\cdot10^{2}$ | $2.64\cdot10^{-1}$ |
| **Auger LMM** | $3.05\cdot10^{3}$ | $1.22\cdot10^{+0}$ | $3.01\cdot10^{3}$ | $1.22\cdot10^{+0}$ | $3.05\cdot10^{3}$ | $1.22\cdot10^{+0}$ | $3.05\cdot10^{3}$ | $1.25\cdot10^{+0}$ |
| **Auger LMX** | $3.68\cdot10^{3}$ | $3.42\cdot10^{-1}$ | $3.63\cdot10^{3}$ | $3.39\cdot10^{-1}$ | $3.68\cdot10^{3}$ | $3.44\cdot10^{-1}$ | $3.67\cdot10^{3}$ | $3.40\cdot10^{-1}$ |
| **Auger LXY** | $4.30\cdot10^{3}$ | $2.33\cdot10^{-2}$ | $4.28\cdot10^{3}$ | $2.24\cdot10^{-2}$ | $4.31\cdot10^{3}$ | $2.39\cdot10^{-2}$ | $4.34\cdot10^{3}$ | $2.11\cdot10^{-2}$ |
| **CK MMX** | $9.39\cdot10^{1}$ | $1.30\cdot10^{+0}$ | $7.58\cdot10^{1}$ | $1.04\cdot10^{+0}$ | $1.00\cdot10^{2}$ | $1.32\cdot10^{+0}$ | $1.27\cdot10^{2}$ | $1.44\cdot10^{+0}$ |
| **Auger MXY** | $4.08\cdot10^{2}$ | $3.22\cdot10^{+0}$ | $3.80\cdot10^{2}$ | $3.24\cdot10^{+0}$ | $4.22\cdot10^{2}$ | $3.23\cdot10^{+0}$ | $4.61\cdot10^{2}$ | $3.28\cdot10^{+0}$ |
| **CK NNX** | $1.99\cdot10^{1}$ | $2.18\cdot10^{+0}$ | $2.74\cdot10^{1}$ | $1.21\cdot10^{+0}$ | $2.19\cdot10^{1}$ | $3.20\cdot10^{+0}$ | $2.99\cdot10^{1}$ | $3.51\cdot10^{+0}$ |
| **Auger NXY** | $2.23\cdot10^{1}$ | $3.15\cdot10^{+0}$ | $4.53\cdot10^{1}$ | $1.40\cdot10^{+0}$ | $1.41\cdot10^{1}$ | $1.02\cdot10^{+1}$ | $3.24\cdot10^{1}$ | $1.09\cdot10^{+1}$ |
| **CK OOX** | – | – | – | – | – | – | $6.00\cdot10^{0}$ | $3.66\cdot10^{+0}$ |
| **X-ray $K_{\alpha}$1** | $2.76\cdot10^{4}$ | $7.41\cdot10^{-1}$ | $2.76\cdot10^{4}$ | $7.62\cdot10^{-1}$ | $2.76\cdot10^{4}$ | $7.41\cdot10^{-1}$ | $2.75\cdot10^{4}$ | $7.51\cdot10^{-1}$ |
| **X-ray $K_{\alpha}$2** | $2.73\cdot10^{4}$ | $3.98\cdot10^{-1}$ | $2.73\cdot10^{4}$ | $4.03\cdot10^{-1}$ | $2.73\cdot10^{4}$ | $4.01\cdot10^{-1}$ | $2.72\cdot10^{4}$ | $3.94\cdot10^{-1}$ |
| **X-ray $K_{\beta}$1** | $3.11\cdot10^{4}$ | $1.31\cdot10^{-1}$ | $3.11\cdot10^{4}$ | $1.35\cdot10^{-1}$ | $3.11\cdot10^{4}$ | $1.34\cdot10^{-1}$ | $3.10\cdot10^{4}$ | $1.38\cdot10^{-1}$ |
| **X-ray $K_{\beta}$2** | $3.18\cdot10^{4}$ | $4.15\cdot10^{-2}$ | $3.18\cdot10^{4}$ | $3.96\cdot10^{-2}$ | $3.18\cdot10^{4}$ | $4.04\cdot10^{-2}$ | $3.17\cdot10^{4}$ | $4.03\cdot10^{-2}$ |
| **X-ray $K_{\beta}$3** | $3.10\cdot10^{4}$ | $6.84\cdot10^{-2}$ | $3.11\cdot10^{4}$ | $6.71\cdot10^{-2}$ | $3.10\cdot10^{4}$ | $6.86\cdot10^{-2}$ | $3.09\cdot10^{4}$ | $6.85\cdot10^{-2}$ |
| **X-ray KM+** | $3.17\cdot10^{4}$ | $3.86\cdot10^{-3}$ | $3.14\cdot10^{4}$ | $1.50\cdot10^{-3}$ | $3.17\cdot10^{4}$ | $3.65\cdot10^{-3}$ | $3.16\cdot10^{4}$ | $4.20\cdot10^{-3}$ |
| **X-ray L** | $3.93\cdot10^{3}$ | $1.50\cdot10^{-1}$ | $3.91\cdot10^{3}$ | $1.61\cdot10^{-1}$ | $3.93\cdot10^{3}$ | $1.49\cdot10^{-1}$ | $2.93\cdot10^{3}$ | $1.32\cdot10^{-1}$ |
| **X-ray M** | $5.46\cdot10^{2}$ | $8.38\cdot10^{-3}$ | – | – | $5.56\cdot10^{2}$ | $7.75\cdot10^{-3}$ | $5.42\cdot10^{2}$ | $4.00\cdot10^{-3}$ |
| **X-ray N** | $9.16\cdot10^{1}$ | $1.95\cdot10^{-1}$ | – | – | $9.91\cdot10^{1}$ | $6.40\cdot10^{-2}$ | – | – |

Table VI: Comparison of calculated Auger and Coster-Kronig yields with the literature for $^{123}I$, $^{124}I$ and $^{125}I$.

| | Isolated atom | | | Condensed phase | | | | |
|---|---|---|---|---|---|---|---|---|
| **Radio-nuclide** | This work | Stepanek (2000) | Pomplun (2000, 2012) | This work | Nikjoo et al. (2008) | Howell (1992) | Stepanek (2000) | Pomplun (2000) |
| **$^{123}I$** | 7.4 | | 7.3[a] | 12.5 | | 14.9 | | |
| **$^{124}I$** | 5.03 | | | 8.43 | 8.2 | | | |
| **$^{125}I$** | 11.9 | 8.92 | 12.2 | 20.2 | 20.2 | 24.9 | 14.5 | 18.3 |

[a] Included 0.9 shake-off electrons.

Table VII: * **Appendix A.** Examples of Monte Carlo calculated spectra of radiations released in independent decays of $^{123}$I in the condensed phase. Coster-Kronig transitions are included in Auger transitions.

| Decay No. | Radiation | No. of rad. | Energy (eV) | | | | | | | |
|---|---|---|---|---|---|---|---|---|---|---|
| **1** | $\gamma$ | 1 | 158970.0 | | | | | | | |
| | X-ray | 1 | 27297.2 | | | | | | | |
| | Auger | 17 | 2829.2 | 135.4 | 388.9 | 84.1 | 432.2 | 3.2 | 8.0 | 16.8 |
| | | | 2.8 | 0.2 | 10.8 | 12.9 | 7.3 | 25.6 | 26.7 | 13.6 |
| | | | 18.5 | | | | | | | |
| **2** | IC | 1 | 127156.2 | | | | | | | |
| | X-ray | 2 | 27571.6 | 27571.7 | | | | | | |
| | Auger | 20 | 3280.7 | 1.7 | 778.3 | 26.7 | 4.9 | 12.1 | 26.7 | 26.7 |
| | | | 13.6 | 17.8 | 2904.9 | 70.1 | 317.7 | 395.5 | 146.5 | 26.1 |
| **3** | IC | 1 | 127156.2 | | | | | | | |
| | X-ray | 2 | 3778.0 | 27297.2 | | | | | | |
| | Auger | 12 | 547.6 | 465.8 | 8.0 | 14.4 | 28.7 | 3412.1 | 475.7 | 2.4 |
| | | | 7.9 | 478.4 | 19.1 | 11.5 | | | | |
| **4** | $\gamma$ | 1 | 158970.0 | | | | | | | |
| | Auger | 5 | 4325.6 | 465.8 | 8.0 | 14.4 | 18.5 | | | |
| **5** | $\gamma$ | 1 | 158970.0 | | | | | | | |
| | X-ray | 1 | 27297.2 | | | | | | | |
| | Auger | 9 | 3692.0 | 16.4 | 151.3 | 476.6 | 20.9 | 13.2 | 26.7 | 13.6 |
| | | | 18.5 | | | | | | | |
| **6** | $\gamma$ | 1 | 158970.0 | | | | | | | |
| | X-ray | 1 | 27297.2 | | | | | | | |
| | Auger | 11 | 3127.7 | 488.2 | 11.4 | 28.2 | 550.9 | 16.9 | 7.7 | 1.5 |
| | | | 83.7 | 25.4 | 7.0 | | | | | |
| **7** | $\gamma$ | 1 | 158970.0 | | | | | | | |
| | X-ray | 1 | 27571.6 | | | | | | | |
| | Auger | 12 | 3376.3 | 91.0 | 314.2 | 2.2 | 92.6 | 8.2 | 19.6 | 10.3 |
| | | | 7.8 | 22.5 | 13.6 | 18.5 | | | | |
| **8** | $\gamma$ | 1 | 158970.0 | | | | | | | |
| | X-ray | 1 | 31098.9 | | | | | | | |
| | Auger | 7 | 99.1 | 17.6 | 11.5 | 463.6 | 8.0 | 26.1 | 19.3 | |
| **9** | $\gamma$ | 1 | 158970.0 | | | | | | | |
| | X-ray | 2 | 27571.6 | 4311.0 | | | | | | |
| | Auger | 1 | 5.4 | | | | | | | |
| **10** | $\gamma$ | 1 | 158970.0 | | | | | | | |
| | X-ray | 1 | 31801.8 | | | | | | | |
| | Auger | 3 | 24.9 | 14.4 | 18.5 | | | | | |
| **11** | $\gamma$ | 1 | 158970.0 | | | | | | | |
| | X-ray | 1 | 31098.9 | | | | | | | |
| | Auger | 9 | 109.8 | 385.7 | 6.8 | 9.9 | 7.5 | 8.9 | 22.3 | 14.4 |
| | | | 18.5 | | | | | | | |
| **12** | $\gamma$ | 1 | 158970.0 | | | | | | | |
| | Auger | 13 | 181.4 | 3399.9 | 472.3 | 5.1 | 11.7 | 17.5 | 11.0 | 392.1 |
| | | | 8.8 | 10.5 | 8.5 | 25.1 | 18.5 | | | |
| **13** | $\gamma$ | 1 | 158970.0 | | | | | | | |
| | X-ray | 1 | 27571.6 | | | | | | | |
| | Auger | 11 | 2915.2 | 395.4 | 23.9 | 60.8 | 18.2 | 217.0 | 406.7 | 16.0 |
| | | | 24.9 | 15.4 | 18.5 | | | | | |
| **14** | IC | 1 | 127156.2 | | | | | | | |
| | X-ray | 1 | 31046.7 | | | | | | | |
| | Auger | 18 | 49.4 | 8.4 | 24.6 | 1.5 | 646.2 | 14.9 | 24.9 | 13.8 |
| | | | 19.3 | 141.2 | 373.5 | 0.5 | 5.8 | 1.8 | 7.3 | 14.9 |
| | | | 7.8 | 11.5 | | | | | | |
| **15** | $\gamma$ | 2 | 528960.0 | 158970.0 | | | | | | |
| | X-ray | 1 | 31791.8 | | | | | | | |
| | Auger | 3 | 34.9 | 13.8 | 20.0 | | | | | |

Table VIII: * **Appendix B.** Examples of Monte Carlo calculated spectra of radiations released in independent decays of $^{124}I$ in the condensed phase. Coster-Kronig transitions are included in Auger transitions.

| Decay No. | Radiation | No. of rad. | Energy (eV) | | | | | | | |
|---|---|---|---|---|---|---|---|---|---|---|
| **1** | $\beta+$ | 1 | | | | | | | | |
| **2** | $\beta+$ | 1 | | | | | | | | |
| **3** | $\gamma$ | 2 | 1690960.0 | 602730.0 | | | | | | |
| | X-ray | 1 | 27297.2 | | | | | | | |
| | Auger | 12 | 3133.2 | 692.4 | 287.2 | 16.3 | 2.8 | 0.2 | 23.6 | 9.9 |
| | | | 23.0 | 59.0 | 13.0 | 30.2 | | | | |
| **4** | $\beta+$ | 1 | | | | | | | | |
| **5** | $\gamma$ | 2 | 722780.0 | 602730.0 | | | | | | |
| | X-ray | 1 | 27571.6 | | | | | | | |
| | Auger | 7 | 3637.2 | 6.8 | 10.6 | 464.2 | 22.3 | 13.8 | 20.0 | |
| **6** | X-ray | 1 | 27571.6 | | | | | | | |
| | Auger | 9 | 3148.0 | 412.1 | 454.0 | 0.8 | 6.3 | 8.9 | 8.4 | 14.4 |
| | | | 18.5 | | | | | | | |
| **7** | $\gamma$ | 2 | 1690960.0 | 602730.0 | | | | | | |
| | Auger | 21 | 22214.7 | 3428.0 | 356.7 | 477.5 | 14.6 | 18.9 | 12.0 | 448.5 |
| | | | 2619.2 | 130.3 | 101.7 | 422.9 | 385.8 | 112.9 | 60.3 | 3.8 |
| | | | 10.6 | 1.6 | 8.3 | 14.3 | 20.0 | | | |
| **8** | Auger | 11 | 2889.6 | 881.3 | 60.0 | 25.9 | 6.9 | 269.8 | 450.9 | 1.3 |
| | | | 8.5 | 13.6 | 17.8 | | | | | |
| **9** | $\beta+$ | 1 | | | | | | | | |
| | $\gamma$ | 1 | 602730.0 | | | | | | | |
| **10** | Auger | 18 | 22797.2 | 3678.2 | 107.0 | 2822.4 | 97.2 | 455.2 | 321.9 | 228.2 |
| | | | 77.1 | 50.7 | 4.5 | 76.8 | 3.9 | 10.6 | 1.5 | 7.2 |
| | | | 14.3 | 30.2 | | | | | | |
| **11** | X-ray | 1 | 27571.6 | | | | | | | |
| | Auger | 11 | 3148.0 | 404.2 | 24.3 | 7.2 | 10.9 | 393.5 | 7.5 | 8.9 |
| | | | 7.2 | 15.1 | 7.0 | | | | | |
| **12** | $\gamma$ | 1 | 602730.0 | | | | | | | |
| | X-ray | 1 | 27571.6 | | | | | | | |
| | Auger | 14 | 2915.2 | 20.0 | 113.9 | 389.3 | 3.2 | 371.2 | 0.1 | 6.8 |
| | | | 9.9 | 7.5 | 16.0 | 24.9 | 14.4 | 18.5 | | |
| **13** | $\gamma$ | 2 | 722780.0 | 602730.0 | | | | | | |
| | X-ray | 2 | 27297.2 | 4041.2 | | | | | | |
| | Auger | 5 | 406.7 | 10.5 | 8.0 | 0.7 | 5.4 | | | |
| **14** | X-ray | 1 | 27571.6 | | | | | | | |
| | Auger | 11 | 3159.5 | 404.0 | 8.5 | 385.7 | 6.8 | 9.9 | 7.5 | 14.9 |
| | | | 26.7 | 26.0 | 5.4 | | | | | |
| **15** | $\beta+$ | 1 | | | | | | | | |
| | IC | 1 | 570916.2 | | | | | | | |
| | X-ray | 1 | 27571.6 | | | | | | | |
| | Auger | 14 | 2653.7 | 77.5 | 273.7 | 22.4 | 412.2 | 12.3 | 6.3 | 0.3 |
| | | | 57.5 | 1.4 | 7.3 | 13.1 | 7.8 | 11.5 | | |

Table IX: * **Appendix C.** Examples of Monte Carlo calculated spectra of radiations released in independent decays of $^{125}$I in the condensed phase. Coster-Kronig transitions are included in Auger transitions.

| **Decay No.** | **Radiation** | **No. of rad.** | **Energy (eV)** | | | | | | | |
|---|---|---|---|---|---|---|---|---|---|---|
| **1** | IC | 1 | 3678.4 | | | | | | | |
| | X-ray | 2 | 31801.8 | 27297.2 | | | | | | |
| | Auger | 8 | 26.7 | 26.0 | 5.4 | 3973.7 | 8.7 | 478.4 | 6.1 | 13.1 |
| **2** | IC | 1 | 3678.4 | | | | | | | |
| | X-ray | 1 | 31099.0 | | | | | | | |
| | Auger | 26 | 23073.0 | 3777.1 | 3102.1 | 482.9 | 269.3 | 321.6 | 0.4 | 2.1 |
| | | | 14.5 | 15.1 | 34.9 | 10.8 | 12.9 | 7.3 | 7.8 | 6.8 |
| | | | 13.8 | 20.0 | 28.8 | 375.5 | 0.3 | 5.7 | 0.5 | 8.5 |
| | | | 107.4 | 0.1 | | | | | | |
| **3** | IC | 1 | 3678.4 | | | | | | | |
| | X-ray | 2 | 147.3 | 27571.7 | | | | | | |
| | Auger | 30 | 23073.0 | 2647.3 | 2433.3 | 765.2 | 749.4 | 605.0 | 95.0 | 391.3 |
| | | | 16.2 | 54.2 | 1.1 | 5.7 | 14.9 | 6.5 | 12.6 | 18.5 |
| | | | 2653.6 | 483.5 | 17.0 | 57.7 | 428.9 | 0.4 | 12.1 | 35.9 |
| | | | 5.6 | 14.5 | 5.4 | 16.8 | 5.9 | 10.8 | | |
| **4** | IC | 1 | 3678.4 | | | | | | | |
| | X-ray | 2 | 27297.2 | 27571.7 | | | | | | |
| | Auger | 18 | 3422.4 | 392.3 | 0.01 | 4.4 | 24.8 | 385.8 | 21.1 | 17.1 |
| | | | 7.2 | 26.8 | 19.3 | 3137.7 | 474.3 | 3.8 | 322.2 | 1.1 |
| | | | 145.6 | 21.6 | | | | | | |
| **5** | IC | 1 | 30553.0 | | | | | | | |
| | X-ray | 1 | 27297.2 | | | | | | | |
| | Auger | 23 | 3176.1 | 72.3 | 530.3 | 11.2 | 385.2 | 7.1 | 9.9 | 8.3 |
| | | | 19.3 | 8.5 | 26.0 | 18.5 | 456.1 | 3443.4 | 1.7 | 164.7 |
| | | | 441.8 | 6.7 | 11.4 | 5.4 | 13.5 | 19.5 | 23.2 | |
| **6** | IC | 1 | 3678.4 | | | | | | | |
| | X-ray | 1 | 27571.7 | | | | | | | |
| | Auger | 20 | 26665.4 | 3376.3 | 408.2 | 466.5 | 13.1 | 3.8 | 459.0 | 1.9 |
| | | | 8.8 | 14.9 | 24.9 | 15.4 | 18.5 | 3149.2 | 473.3 | 438.9 |
| | | | 7.3 | 1.6 | 5.0 | 10.8 | | | | |
| **7** | IC | 1 | 3678.4 | | | | | | | |
| | X-ray | 2 | 3778.1 | 27571.7 | | | | | | |
| | Auger | 17 | 26763.6 | 107.0 | 532.2 | 11.9 | 15.1 | 477.6 | 15.4 | 17.8 |
| | | | 3137.7 | 482.8 | 3.1 | 363.7 | 0.5 | 5.6 | 25.0 | 6.1 |
| | | | 0.1 | | | | | | | |
| **8** | IC | 1 | 3678.4 | | | | | | | |
| | X-ray | 2 | 27297.2 | 31099.0 | | | | | | |
| | Auger | 27 | 2886.4 | 22.7 | 18.0 | 470.4 | 4.2 | 10.9 | 16.4 | 105.8 |
| | | | 4.1 | 380.8 | 16.3 | 7.1 | 9.9 | 8.9 | 13.2 | 34.9 |
| | | | 15.4 | 18.5 | 88.8 | 6.3 | 358.0 | 6.8 | 11.8 | 8.3 |
| | | | 14.9 | 6.1 | 23.2 | | | | | |
| **9** | IC | 1 | 3678.4 | | | | | | | |
| | X-ray | 2 | 27297.2 | 27571.7 | | | | | | |
| | Auger | 17 | 3983.7 | 15.6 | 488.3 | 26.1 | 7.0 | 3137.7 | 400.9 | 299.9 |
| | | | 9.1 | 12.1 | 16.6 | 8.4 | 11.8 | 8.3 | 11.8 | 8.3 |
| | | | 16.8 | 5.9 | 11.5 | | | | | |
| **10** | $\gamma$ | 1 | 35492.2 | | | | | | | |
| | Auger | 1 | 3983.7 | | | | | | | |
| **11** | IC | 1 | 3678.4 | | | | | | | |
| | X-ray | 2 | 27571.6 | 27571.7 | | | | | | |
| | Auger | 22 | 3159.5 | 486.9 | 215.5 | 80.2 | 2.0 | 8.7 | 13.9 | 33.6 |
| | | | 14.3 | 20.0 | 3377.4 | 20.6 | 4.4 | 442.5 | 7.5 | 13.4 |
| | | | 8.4 | 12.7 | 8.3 | 14.9 | 7.8 | 11.5 | | |
| **12** | IC | 1 | 3678.4 | | | | | | | |
| | X-ray | 2 | 31801.8 | 27571.7 | | | | | | |
| | Auger | 12 | 26.7 | 26.0 | 17.8 | 3407.5 | 20.8 | 130.8 | 466.9 | 13.1 |
| | | | 5.6 | 25.0 | 6.8 | 11.5 | | | | |
| **13** | $\gamma$ | 1 | 35492.2 | | | | | | | |
| | Auger | 13 | 549.6 | 16.7 | 3159.5 | 486.9 | 10.4 | 244.3 | 8.1 | 35.8 |
| | | | 7.3 | 22.9 | 14.9 | 69.5 | 7.0 | | | |
| **14** | IC | 1 | 3678.4 | | | | | | | |
| | X-ray | 3 | 27571.6 | 3777.5 | 27571.7 | | | | | |
| | Auger | 12 | 479.4 | 13.6 | 17.8 | 3137.7 | 400.9 | 4.9 | 457.7 | 12.6 |
| | | | 4.5 | 14.9 | 7.8 | 10.8 | | | | |
| **15** | IC | 1 | 3678.4 | | | | | | | |
| | X-ray | 2 | 27571.6 | 27279.2 | | | | | | |
| | Auger | 34 | 2663.8 | 77.1 | 383.5 | 51.3 | 11.2 | 441.7 | 34.0 | 11.9 |
| | | | 0.1 | 20.8 | 9.9 | 7.5 | 10.5 | 9.8 | 13.6 | 18.5 |
| | | | 2876.2 | 129.4 | 123.5 | 12.4 | 290.7 | 288.0 | 0.8 | 16.1 |
| | | | 0.4 | 7.7 | 26.3 | 0.5 | 8.4 | 2.1 | 17.6 | 13.5 |
| | | | 7.5 | 0.1 | | | | | | |